\documentclass[a4paper,fleqn,usenatbib]{mnras}
\pdfoutput=1

\usepackage[T1]{fontenc}
\usepackage{ae,aecompl}
\usepackage{epstopdf}
\usepackage{aas_macros}
\usepackage{braket}
\usepackage{verbatim}
\usepackage{wrapfig}
\usepackage{booktabs}
\usepackage{amsfonts}
\usepackage{amsmath}
\usepackage{appendix}
\usepackage{graphicx}
\usepackage[usenames,dvipsnames]{color}
\usepackage[normalem]{ulem}
\usepackage{array}
\title[Accounting for sample selection in Bayesian analyses]{Accounting for sample selection in Bayesian analyses}

\author[S. R. Hinton et al.]{Samuel R. Hinton,$^{1,2}$\thanks{E-mail: samuelreay@gmail.com}
	Alex G. Kim,$^{3}$
	Tamara M. Davis$^{1,2}$
\\
$^{1}$School of Mathematics and Physics, The University of Queensland, Brisbane, QLD 4072, Australia\\
$^{2}$ARC Centre of Excellence for All-sky Astrophysics (CAASTRO)\\
$^{3}$Physics Division, Lawrence Berkeley National Laboratory, 1 Cyclotron Road, Berkeley, CA 94720, USA
}
\date{Accepted XXX. Received YYY; in original form ZZZ}

\pubyear{2017}

\begin{document}

\label{firstpage}
\pagerange{\pageref{firstpage}--\pageref{lastpage}}
\maketitle

\begin{abstract}
Astronomers are often confronted with funky populations and distributions of objects: brighter objects are more likely to be detected; targets are selected based on colour cuts; imperfect classification yields impure samples. Failing to account for these effects leads to biased analyses. In this paper we present a simple overview of a Bayesian consideration of sample selection, giving solutions to both analytically tractable and intractable models. This is accomplished via a combination of analytic approximations and Monte Carlo integration, in which dataset simulation is efficiently used to correct for issues in the observed dataset. This methodology is also applicable for data truncation, such as requiring densities to be strictly positive. Toy models are included for demonstration, along with discussions of numerical considerations and how to optimise for implementation.  We provide sample code to demonstrate the techniques.  The methods in this paper should be widely applicable in fields beyond astronomy, wherever sample selection effects occur.
\end{abstract}

\nokeywords

\section{Introduction}

Sample selection is a problem in many areas of scientific inquiry. For example, it is one of the primary difficulties when performing supernova cosmology analysis, as telescopes more easily detect bright objects than faint ones, causing our observed supernova distribution to deviate from the actual underlying distribution. This  ``Malmquist Bias'', has been a subject of much investigation \citep{Butkevich2005}. It is taken into account during analyses by either modifying the observed data to remove the expected bias \citep{BetouleKessler2014, ConleyGuySullivan2011}, or by incorporating the expected bias into the underlying model \citep{Rubin2015}. Selection effects feature prominently in most fields of astronomy, and are critical in investigations of gravitational waves \citep{MessengerVeitch2013, AbbottAbbott2016}, luminosity functions \citep{MarshallTananbaum1983, KellyFan2008, BuchnerGeorgakakis2015}, trans-Neptunian objects \citep{Loredo2004} and much more.  Truncated data are also commonly encountered in biological fields, where data such as mortality rates are left-truncated \citep{JANE1898}. Simplified and generalised examples have been investigated in numerous fashions \citep{woodroofe1985estimating, Gull1989bayesian, grogger1991models, o1995truncated} and with application to specific fitting algorithms \citep{Gelfand1992, Liang2010}. Whilst generalised resources exist that provide a comprehensive overview of sample selection and analysis techniques in a similar fashion to this work \citep{klein2005survival, andreon2015bayesian}, these sources are often opaque due to volume and mathematical complexity. 

This work provides a simple treatment of sample selection using a common Bayesian technique. The general theory for considering selection effects is discussed in Section \ref{sec:theory}. Section \ref{sec:examples} provides three examples of increasing complexity with sample selection. Section \ref{sec:tricks} details numeric concerns and tricks to be aware of for effective implementation of Monte Carlo corrections applied to analytic approximations.

\section{Theory}
\label{sec:theory}

When formulating and fitting a model using a constraining dataset, we wish to resolve the posterior surface defined by
\begin{align}
P(\theta | {\rm data}) \propto P({\rm data} | \theta) P(\theta),
\end{align}
which gives the probability of the model parameter values ($\theta$) given the data.  Prior knowledge of the allowed values of the model parameters is encapsulated in the prior probability $P(\theta)$. Of primary interest to us is the likelihood of observing the data given our parametrised model, $\mathcal{L} \equiv P({\rm data} | \theta)$. When dealing with experiments which have imperfect selection efficiency, our likelihood must take that efficiency into account.  We need to describe the probability that the events we observe are both drawn from the distribution predicted by the underlying theoretical model \textit{and} that those events, given they happened, are subsequently successfully observed.  To make this extra conditional explicit, we write the likelihood of the data given an underlying model, $\theta$, \textit{and} that the data are included in our sample, denoted by $S$, as:
\begin{align}
\mathcal{L} &= P({\rm data} | \theta, S). \label{eq:like}
\end{align}
A variety of selection criteria are possible, and in our method we use our data in combination with the proposed model to determine the probability of particular selection criteria.  That is, we characterise a function $P(S|{\rm data},\theta)$, which colloquially can be stated as \textit{the probability of a potential observation passing selection cuts, given our observations and the underlying model}. We can introduce this expression in a few lines due to symmetry of joint probabilities and utilising that $P(x,y,z) = P(x|y,z)P(y,z) = P(y|x, z)P(x, z)$:
\begin{align}
P({\rm data} | S , \theta) P(S, \theta) &= P(S | {\rm data}, \theta) P({\rm data}, \theta)\\
P({\rm data} | S , \theta) &= \frac{ P(S | {\rm data}, \theta) P({\rm data}, \theta) }{ P(S, \theta) }\\
 &= \frac{ P(S | {\rm data}, \theta) P({\rm data} | \theta) P(\theta)}{ P(S | \theta)  P(\theta)}\\
 &= \frac{ P(S | {\rm data}, \theta) P({\rm data} | \theta)}{ P(S | \theta) }
\end{align}
which is equal to the likelihood $\mathcal{L}$. Introducing an integral over all possible events $D$, so we can evaluate $P(S|\theta)$, 
\begin{align}
\mathcal{L}&= \frac{P(S|{\rm data},\theta) P({\rm data}|\theta)}{\int P(S, D|\theta)\, dD} \\
\mathcal{L} &= \frac{P(S|{\rm data},\theta) P({\rm data}|\theta)}{\int P(S | D, \theta) P(D|\theta)\, dD}, \label{eq:main}
\end{align}
where the integral in the denominator has the same dimensionality as the experimental data. \textbf{Equation \ref{eq:main} is the generalised likelihood of experiments with sample selection.}

\section{Sample Selection}
\label{sec:examples}

In this Section we present sample selections of increasing complexity. Each sample selection is accompanied by an illustrative example inspired by Type Ia supernova cosmology, where we characterize the properties of a standard or standardizable candle. Python code implementations of all examples, including plot generation, can be found at the Github repository \verb|SampleSelection|.\footnote{\url{https://github.com/samreay/SampleSelection}}
 
Section \ref{sec:real} details the case of an analytically intractable posterior, which we solve using a combination of analytic approximation and Monte-Carlo integration. For \textit{doubly-intractable} likelihoods where the denominator has no closed form,  the reader is recommended auxiliary variable methods \citep{MollerPettitt2006, MurrayGhahramani2012, Liang2010}.
 
\subsection{Complete Selection}
\label{sec:perfect}
In a perfect world, data is neither biased nor truncated. Data is perfect. Uncertainty does not exist. Presumably everything is also spherical and in a vacuum.

\begin{figure}
	\begin{center}
		\includegraphics[width=\columnwidth]{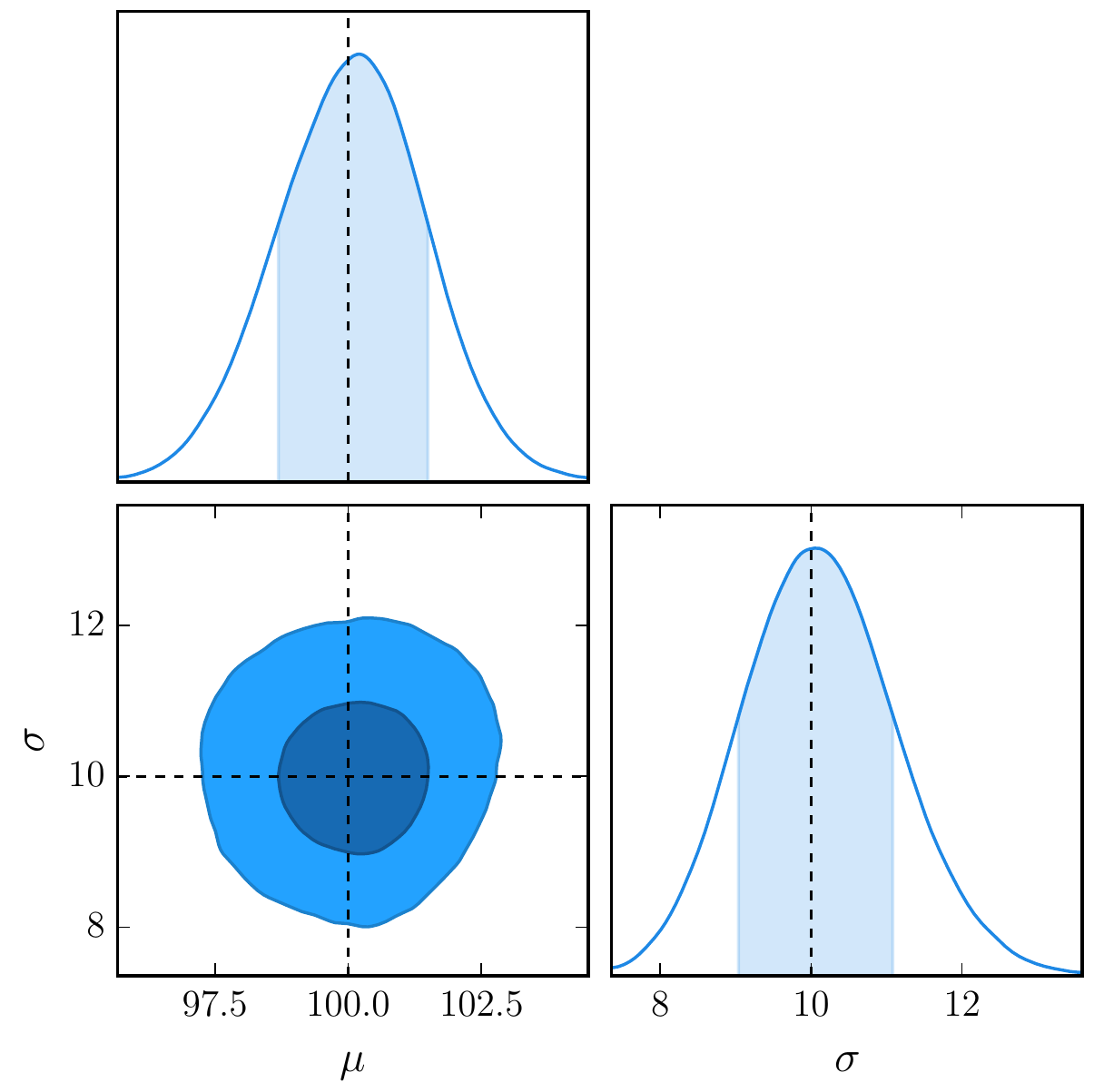}
	\end{center}
	\caption{A systematic test of the model in Section \ref{sec:perfect}, done by stacking the output chains from fitting 100 independent realisations of our 100 data points. Any systematic offset in our model would be revealed by a shift in the stacked results away from the true parameter values. This plot can be generated by executing  \mbox{\texttt{0\_perfect.py}}}.
	\label{fig:perfect}
\end{figure}

We thus begin by considering an ideal situation where the sample is complete.  All events are included in the sample such that $P(S | {\rm data}, \theta)=1$.  Trivially this expression is independent of our data and model parameters $\theta$, and our likelihood from Equation \eqref{eq:main} reduces down to Equation \eqref{eq:like}. As a concrete example of this case, let us consider a model for a population of objects whose brightnesses form a normal distribution with average $\mu$ and standard deviation $\sigma$. Let us also assume that our experiment produces data $x_i$, which are independent measurements of unique objects' brightnesses (with negligible measurement uncertainty), then our observations are drawn from a Normal distribution, notated as
\begin{align}
\vec{x} \sim \mathcal{N}(\mu,\sigma).
\end{align}
If, having collected our observations $\vec{x}$, we wanted to constrain $\mu$ and $\sigma$, this would be a simple task of modelling the posterior surface. We simply wish to map the surface
\begin{align}
P(\theta | {\rm data}) \propto P({\rm data} | \theta) P(\theta),
\end{align}
where our model parameters are $\theta = \lbrace \mu, \sigma \rbrace$, giving
\begin{align}
P(\mu,\sigma| \vec{x}) &\propto P(\vec{x} | \mu, \sigma) P(\mu, \sigma).
\end{align}
With uniform priors, $P(\mu,\sigma) = {\rm constant}$, and can be absorbed into the constant of proportionality. Expanding our observation vector to take into account all observations and not just one, the posterior surface is given by
\begin{align}
P(\mu,\sigma| \vec{x}) &\propto \prod_{i=1}^N \mathcal{N}(x_i | \mu, \sigma), \label{eq:prod}
\end{align}
where $\mathcal{N}(x_i|\mu,\sigma)$ represents the probability of drawing sample $x_i$ from the Normal distribution with mean $\mu$ and standard deviation $\sigma$. Generating a hundred data points with $\mu=100,\ \sigma=10$ as an example, we can recover our input parameters easily, as shown in Figure \ref{fig:perfect}.

\subsection{Analytic Sample Selection}
\label{sec:imperfect}

We now consider a case with selection bias that allows us to form an analytic expression for the likelihood.  This illustrative case is useful for the reader as the influence of selection bias is simple and intuitive. The case we consider is identical to the one in the previous section, \textit{except} that only the subset of objects brighter than a threshold $\alpha$ are included in the sample. This case is motivated by instrumental efficiency, however we note that it is mathematically identical to common physical data truncation scenarios --- if we were observing mass, which must be positive, we could incorporate this truncation by setting $\alpha = 0$.

With this instrumental sample selection example, all events satisfy $x_i > \alpha$, giving $P(S|x,\theta) = \mathcal{H}(x - \alpha)$, where $\mathcal{H}$ is the Heaviside step function
\begin{align}
\mathcal{H}(y)\equiv \begin{cases}
1 \quad {\rm if }\  y \ge 0 \\
0 \quad {\rm otherwise.}
\end{cases}
\end{align} 
If we do not take this truncation into account, we will recover biased parameter estimates. However, we can correct for this truncation using Equation \eqref{eq:main}, as the integral in the denominator has an analytic solution. Having successfully observed $x_i$, it follows that $x_i > \alpha$ and so $P(S|x_i,\theta) = 1$. To substitute in our normal model,
\begin{align}
\mathcal{L}_i &= \frac{P(S|x_i,\theta) P(x_i|\theta)}{\int P(S | D, \theta) P(D|\theta)\, dD} \\
&= \frac{ \mathcal{N}(x_i|\mu, \sigma)}{\int_{-\infty}^\infty \mathcal{H}(D - \alpha) \mathcal{N}(D|\mu, \sigma)\, dD} \\
&= \frac{ \mathcal{N}(x_i|\mu, \sigma)}{\int_{\alpha}^\infty \mathcal{N}(D|\mu, \sigma)\, dD} \\
&= \frac{ \mathcal{N}(x_i|\mu, \sigma)}{\frac{1}{2} {\rm erfc}\left[ \frac{\alpha - \mu}{\sqrt{2}\sigma} \right]}, 
\end{align}
where in the last line we have evaluated the integral in the case $\mu > \alpha$. Note that this is for a single observation, and so for a set of independent observations we need to introduce the product found in Equation \eqref{eq:prod}. 
\begin{align}
\mathcal{L} &= \prod_{i=1}^N \mathcal{L}_i =  \prod_{i=1}^N \frac{ \mathcal{N}(x|\mu, \sigma)}{\frac{1}{2} {\rm erfc}\left[ \frac{\alpha - \mu}{\sqrt{2}\sigma} \right]}, 
\end{align}
However, as our selection efficiency correction is independent of our observations $\vec{x}$, we can take it outside the product.
\begin{align}
\mathcal{L} &= 2^N \left( {\rm erfc}\left[ \frac{\alpha - \mu}{\sqrt{2}\sigma} \right] \right)^{-N}\prod_{i=1}^N  \mathcal{N}(x|\mu, \sigma),
\end{align}
We can add this correction to our model, and note that we now recover unbiased parameter estimates. 

\begin{figure}
	\begin{center}
		\includegraphics[width=\columnwidth]{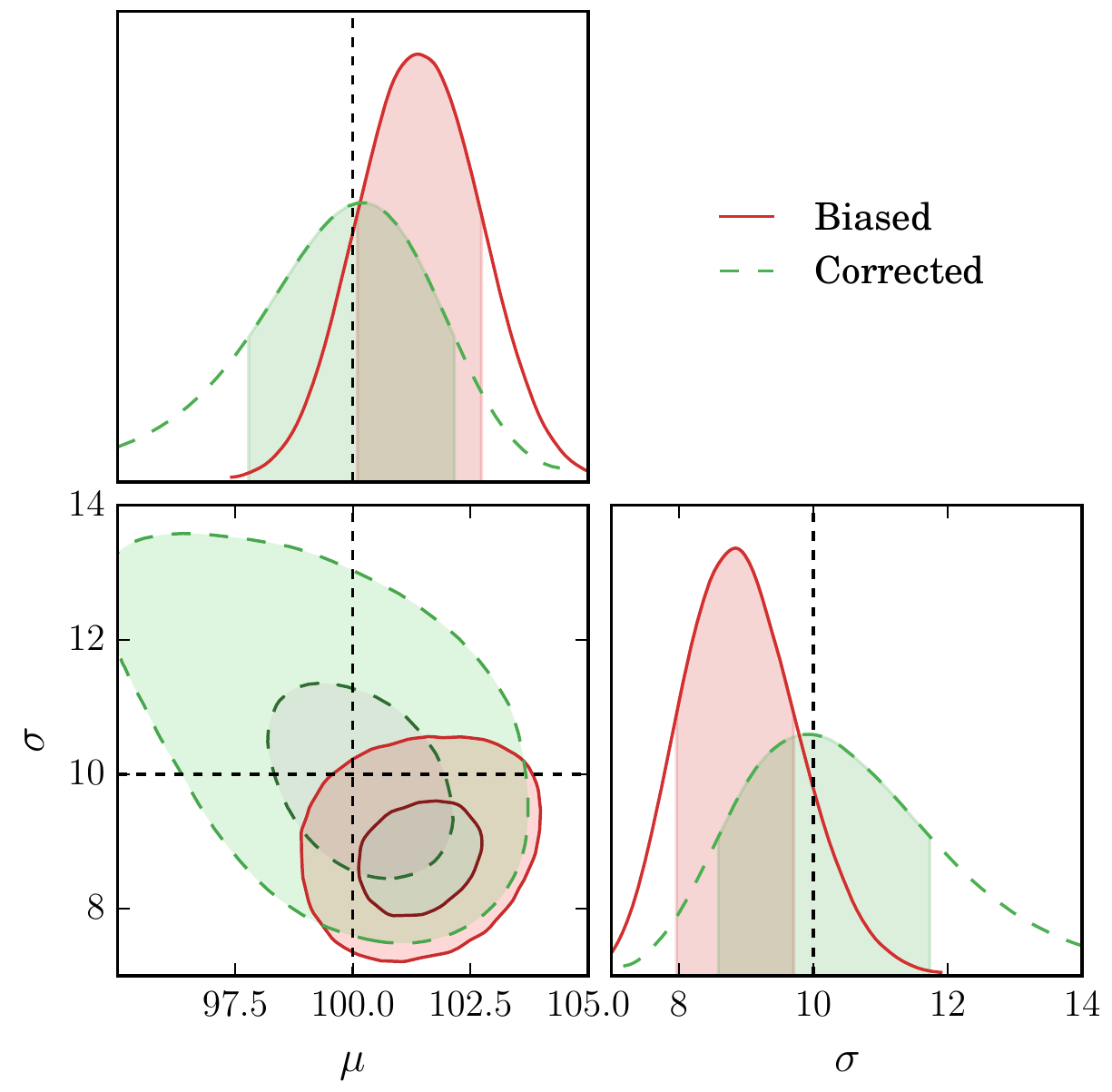}
	\end{center}
	\caption{A systematic test of the model in Section \ref{sec:imperfect}, done by stacking the output chains from fitting 100 independent realisations of our 100 data points, subject to our thresholding. The bias shown in the red `Biased' contour can be corrected to via the techniques shown in Section \ref{sec:imperfect} to recover unbiased surfaces. This plot can be generated by executing  \mbox{\texttt{1\_imperfect.py}}.}
	\label{fig:imperfect}
\end{figure}

Continuing our example introduced in Section \ref{sec:perfect}, we assign an arbitrary value $\alpha=85$, and again fit realisations of $100$ observations.  This is demonstrated in Figure \ref{fig:imperfect}, which shows the posterior surfaces for when you take sample selection into account and when you do not. The selection bias preferentially selects intrinsically brighter objects and, by cutting out some of the distribution, narrows the observed distribution. As such, we note that the bias correction correctly increases the weight of low $\mu$ and high $\sigma$ parametrisations, as those models would be subject to the most objects lost from the sample selection criterion. We also note that not only does the best fit location for each parameter shift, but the shape of the posterior surface itself is significantly modified.

\subsection{Analytically Intractable Sample Selection}
\label{sec:real}
Unfortunately it is a rare scenario when dealing with nature and all her faults for us to have an analytic selection function, let alone a function encapsulated by a single parameter. A more realistic selection efficiency would take the form of non-analytic function of many model parameters. And the function would probably be stochastic too, just to make things more difficult. 

In this subsection we introduce an example that presents computational challenges in the evaluation of the posterior, and which better reflects the model complexity in supernova cosmology analysis. The specific challenge is that the denominator in Equation \eqref{eq:main} is dependent on $\theta$ and does not have an analytic solution (i.e. has to be solved numerically).

A simple solution to analytic intractability in the denominator of Equation \eqref{eq:main} is to only sample from the numerator, adding corrections from the denominator after the fact. However, this strategy rarely works, as ignoring the denominator completely can lead to fits in completely wrong areas of parameter space. A solution is to find an approximate, analytic correction we can utilise in our fitting algorithm which seeks to shift the region of parameter space sampled by the sampler closer to the correct area, and then importance sample our sampled Monte Carlo chains to provide a fully corrected surface. Given an analytic approximation to the denominator $w_{\rm approx}$, we explicitly break our likelihood per observation into two parts, $\mathcal{L}_i = \mathcal{L}_{i1} \mathcal{L}_{i2}$, with the parts given in general by
\begin{align}
\mathcal{L}_{i1} &= \frac{P(S|{\rm data},\theta) P({\rm data}|\theta)}{w_{\rm approx}} \\
\mathcal{L}_{i2} &= \frac{w_{\rm approx}}{\int P(S, D|\theta)\, dD}.
\end{align}
By adding the approximate correction $w_{\rm approx}$ we seek to sample the correct region of parameter space, and provided a method of forward modelling or simulating observations, we can evaluate $\mathcal{L}_{i2}$ numerically using Monte Carlo integration. Whilst other integration algorithms can be used, Monte Carlo integration is an appropriate choice, as it is based on drawing random numbers from underlying distributions and is efficient for evaluating high dimensional integrals. For selection effects which can be encapsulated in only a few dimensions, other numeric integration techniques (such as gridded integration) are effective \citep{BuchnerGeorgakakis2015}. To perform the numeric integration using Monte Carlo integration, we must be able to forward model at each spot in parameter space --- focusing solely on forward modelling would give the approach of Approximate Bayesian Computation, a technique which is growing quickly in popularity in astronomy \citep{CameronPettitt2012, JenningsMadigan2017}.

Let us modify our imperfect toy model from the previous section. Instead of observing just one variable, $x$, we also observe a new independent variable, $y$, which is drawn from its own distribution $y \sim \mathcal{N}(\mu_y, \sigma_y)$, and has no measurement uncertainty (like $x$). Our selection efficiency can now become a combination of $x$ and $y$, such that we only observe events that satisfy $x + \beta y > \alpha$, giving $P(S|x,y,\theta) = \mathcal{H}(x + \beta y - \alpha)$, a relationship reminiscent of the stretch and colour corrections for standardising supernovae.  Our likelihood for such a toy model becomes now the combination of probabilities for observing both $x$ and $y$, with the denominator becoming an integral over all possible $X$ and $Y$ events subject to our selection effects.
\begin{align}
\mathcal{L}_i &= \frac{P(S|x_i, y_i,\theta) P(x_i, y_i|\theta)}{\iint P(S | X, Y, \theta) P(X, Y|\theta)\, dX\,dY},
\end{align}
where $\theta = \lbrace \mu, \sigma, \mu_y, \sigma_y \rbrace$. As before, given a successful observation and a deterministic selection efficiency, $P(S|x_i, y_i,\theta) = 1$. We continue to use uniform priors to keep the example simple. Substituting in the step function selection efficiency for $P(S|X, Y,\theta)$ and the normal distributions for $P(x_i, y_i|\theta)$, we have
\begin{align}
\mathcal{L}_i &= \frac{ \mathcal{N}(x_i|\mu, \sigma) \mathcal{N}(y_i|\mu_y, \sigma_y)}
{\iint_{-\infty}^\infty \mathcal{H}(X + \beta Y - \alpha) \mathcal{N}(X|\mu, \sigma) \mathcal{N}(Y|\mu_y, \sigma_y)\, dX dY}. \label{eq:above}
\end{align}
While we could work directly with the likelihood in Equation \eqref{eq:above}, it is beneficial to seek an analytic approximation. In our example, if $\beta \ll 1$, such that the majority of selection effect is encapsulated by $x$ and not $y$, our approximate correction can take the form found in the previous correction from Section \ref{sec:imperfect}. Having a known $\beta$ and a guess $\mu_{y, {\rm approx}}$, we add in a small adjustment to the analytic correction from Section \ref{sec:imperfect} to account for the expected mean contribution of $y$ into the selection effect:
\begin{align}
w_{\rm approx} = \frac{1}{2} {\rm erfc}\left[ \frac{\alpha - \mu - \beta \mu_{y, {\rm approx}}}{\sqrt{2}\sigma} \right].
\end{align}
This gives our likelihood parts as
\begin{align}
\mathcal{L}_{i1} &= \frac{ \mathcal{N}(x_i|\mu, \sigma) \mathcal{N}(y_i|\mu_y, \sigma_y)}{w_{\rm approx} } \\
\mathcal{L}_{i2} &= \frac{ w_{\rm approx}  }{\iint_{-\infty}^\infty \mathcal{H}(x + \beta y - \alpha) \mathcal{N}(X|\mu, \sigma) \mathcal{N}(Y|\mu_y, \sigma_y)\, dX dY}.
\end{align}
Evaluating $\mathcal{L}_{i2}$ using Monte Carlo integration of $n$ samples, we have for a single observation that
\begin{align}
\mathcal{L}_{i2} = \frac{ w_{\rm approx}  n }{\sum_{j=1}^{n} \mathcal{H}(X_j + \beta Y_j - \alpha) \mathcal{N}(X_j|\mu, \sigma) \mathcal{N}(Y_j|\mu_y, \sigma_y)} \label{eq:mc}.
\end{align}
We now wish to move to a set of observations. As $\mathcal{L}_{i2}$ is independent of the specific observation $i$, this is a simple step:
\begin{align}
\mathcal{L}_1 &= w_{\rm approx}^{-N} \prod_{i=1}^N \mathcal{N}(x_i|\mu, \sigma) \mathcal{N}(y_i|\mu_y, \sigma_y) \\
\mathcal{L}_{2} &= \left(  \frac{ w_{\rm approx}  n }{\sum_{j=1}^{n} \mathcal{H}(X_j + \beta Y_j - \alpha) \mathcal{N}(X_j|\mu, \sigma) \mathcal{N}(Y_j|\mu_y, \sigma_y)}\right)^N
\end{align}
$\mathcal{L}_1$ can thus be fitted with a traditional sampler without numeric difficulty or slowdown, and $\mathcal{L}_2$ allows us to calculate the weight of each sample given by fitting $\mathcal{L}_{1}$. We are effectively importance sampling our likelihood evaluations. The computational benefits of this should not be understated either --- each sample in our chains can be reweighted independently, providing a task that is trivially parallelisable. 

\begin{figure}
	\begin{center}
		\includegraphics[width=\columnwidth]{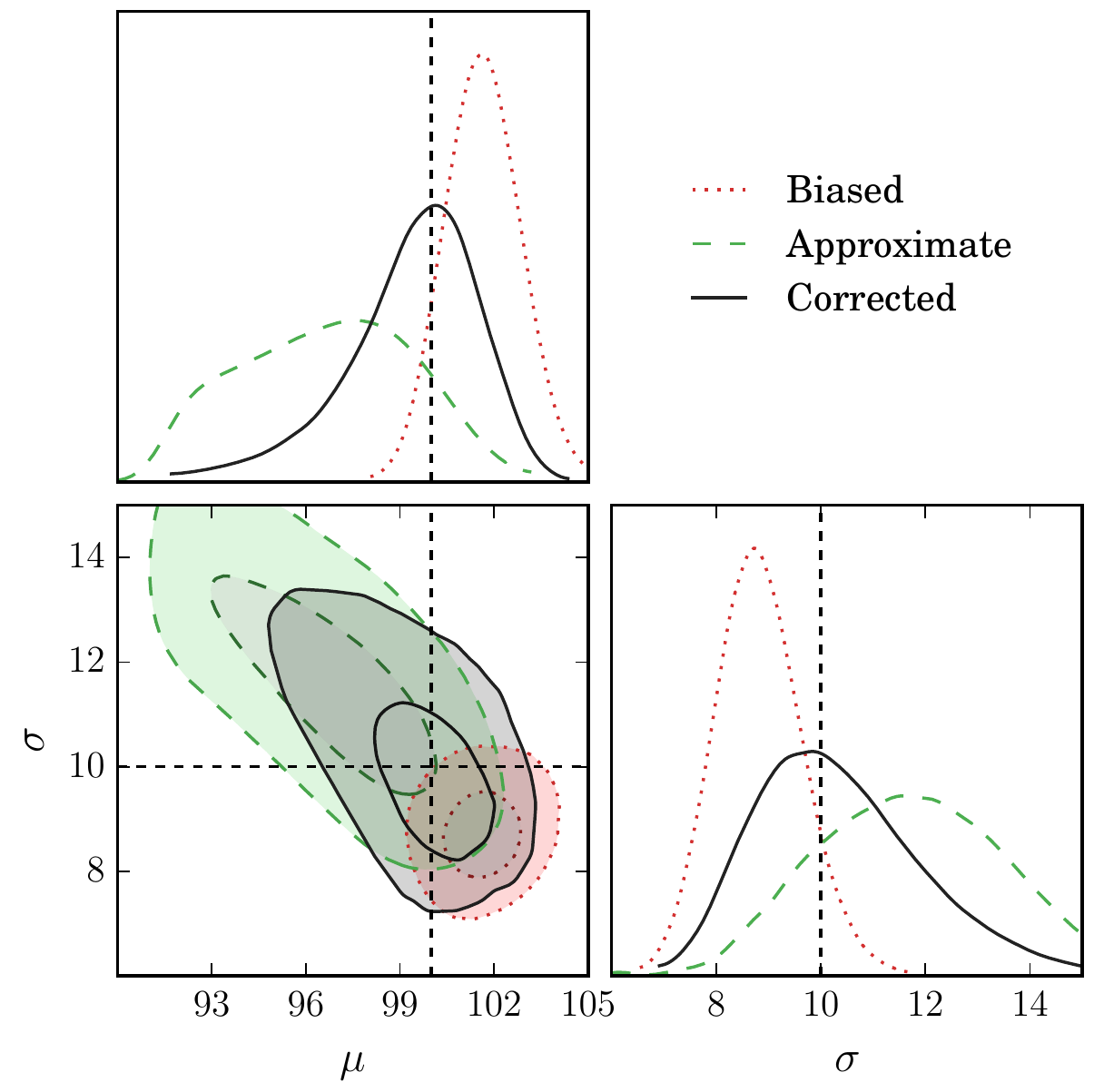}
	\end{center}
	\caption{A systematic test of the model from Section \ref{sec:real}, done by stacking the output chains from fitting 100 independent realisations of our 100 data points, subject to our thresholding. The likelihood $\mathcal{L}_1$ was evaluated, and reweighted using Monte Carlo integration of a hundred thousand possible events as per $\mathcal{L}_2$. The truncated data with no correction is shown as `Biased' in dotted red, the `Approximate` only correction ($\mathcal{L}_1$) shown in dashed green, and the final reweighted chain shown in solid black as `Corrected'. A large number of samples of $\mathcal{L}_1$ had to be generated to ensure sufficient sampling of our `Corrected' posterior. This plot can be generated by executing \mbox{\texttt{2\_real.py}}.}
	\label{fig:real}
\end{figure}

Taking our example, and setting values for our model  $\mu = 100$, $\sigma = 10$, $\mu_y = 30$, $\sigma_y = 5$, $\alpha=92$ and a known $\beta = 0.2$ with $\mu_{y, {\rm approx}} = 20$, we can realise $100$ observations as was done in previous examples, and fit them. We end up with a corrected posterior surface as shown in Figure \ref{fig:real}. It is important to note that the point of maximum likelihood is biased by roughly the same amount in the biased and approximate posterior surfaces in Figure \ref{fig:real}. However, as the approximately correctly posterior is broader than the biased surface, it has far more samples in the region of parameter space mapped by the fully corrected posterior. This is the entire purpose of the approximate correction --- to maximise the number of samples in the correct region of parameter space, so that we can importance sample our chains efficiently.

\section{Numerical Techniques}
\label{sec:tricks}

\subsection{Importance Sampling}
Further tricks can be used to increase the efficiency with which the samples are reweighted. Firstly, the overarching analytic model often provides functions which can be drawn from efficiently. In the case of our example, by drawing the random numbers $X$ and $Y$ respectively from the normal distributions $\mathcal{N}(\mu,\sigma)$ and $\mathcal{N}(\mu_y,\sigma_y)$ (i.e. traditional importance sampling) we need only evaluate the step function for our data points. That is, for a simplified 1D example only in $X$, we replace
\begin{align}
\int f(X) \mathcal{N}(X|\mu,\sigma)\, dX = \frac{1}{N} \sum_{i=1}^N f(X_i) \mathcal{N}(X_i | \mu, \sigma),
\end{align}
where $X$ is drawn from a uniform distribution over all space, with 
\begin{align}
\int f(X) \mathcal{N}(X|\mu,\sigma)\, dX = \frac{1}{N} \sum_{i=1}^N f(X_i),
\end{align}
where $X$ drawn from $\mathcal{N}(X_i | \mu, \sigma)$.

\subsection{Precomputing selection}

The integral in the denominator of Equation \eqref{eq:main} must be recalculated for each set of $\theta$ parameter-values that are considered. A significant cost saving is possible when Monte Carlo integration is done, if the same Monte Carlo realizations are used for all sets of $\theta$. This is especially true if evaluating the sample selection $P(S|{\rm data}, \theta)$ is numerically expensive, but can only be used when the sample selection is independent of $\theta$, so that $P(S|{\rm data}, \theta) = P(S|{\rm data})$, and when there is some prior knowledge of $\theta$ such that we can estimate a reasonable $\theta_{\rm approx}$. Considering the integral from Equation \eqref{eq:main}, we have
\begin{align}
&\int P(S|D, \theta) P(D|\theta) \, dD = \notag \\
&\quad\quad\quad\int P(S|D) \frac{P(D|\theta)}{P(D|\theta_{\rm approx})} \left[ P(D|\theta_{\rm approx}) \,dD \right]
\end{align}
With the mathematics laid as so, it is easy to pregenerate a set of events $D_i$ drawn from distribution $ P(D|\theta_{\rm approx})$, evaluate $P(S|D_i)$, and reuse them for all samples. Formulated using Monte Carlo integration,
\begin{align}
&\int P(S|D, \theta) P(D|\theta) \, dD = \frac{1}{N} \sum_{i=1}^N \frac{P(S|D_i)}{P(D_i|\theta_{\rm approx})} P(D_i|\theta).
\end{align}
The \textit{only} term that needs to be evaluated for each sample is $P(D_i|\theta)$, as all other terms can be pre-computed and stored. When using this technique, take care that the number of events used when calculating the weights is sufficient to make the statistical error of Monte Carlo integration insignificant when compared to the constraining power of your dataset.

Consider the imperfect example --- where we observe $x$ drawn from an underlying normal distribution as was done in Section \ref{sec:imperfect}, but utilise the Monte Carlo integration technique from Section \ref{sec:real}, and work without an analytic approximation (i.e. we set $w_{\rm approx} =1$). We could estimate, given some prior knowledge, that variable $x \approx \mathcal{N}(\mu_{\rm approx}, \sigma_{\rm approx})$. In this example, it would be better to overestimate $\sigma_{\rm approx}$ rather than underestimate it, so that we still effectively sample the correct area of parameter space. We then draw samples of $x$ from this approximate distribution, recording the probability of each draw and then calculating whether our potential observation of $x$ would be observed in the experiment or not. That is, we assign $P(S|x,\theta) = P(S|x) = 1$ or $0$ given it passed cuts or not. We discard all events with $0$ weight (as they have $0$ weight), and only track those events which pass. Then, when calculating the sample reweighting after running chains, $\mathcal{L}_{i2}$ becomes
\begin{align}
\mathcal{L}_{i2} \propto w_{\rm approx}  \left[\sum_{j=1}^{n} \frac{\mathcal{N}(X_j | \mu, \sigma)}{\mathcal{N}(X_j | \mu_{\rm approx}, \sigma_{\rm approx})} \right]^{-1},
\end{align} 
where you can see that we discard the constant $n$ from Equation \eqref{eq:mc} as we only care about likelihood proportionality. Provided our parameter estimate is reasonably well informed, the computation benefit this precomputation provides is enormous for any nontrivial selection function. Not only do we now waste no time when calculating $\mathcal{L}_{i2}$ determining $P(S|{\rm data}, \theta)$, as we only save results that pass the cuts, we have no wasted evaluations of $\mathcal{N}(X_j | \mu, \sigma)$.

As stated previously, this technique requires that selection efficiency of an observation is independent of model parameters $\theta$. For many experimental cases this may hold, however if it does not this method cannot be used to increase efficiency. Gridding or interpolating the parameter space is strongly not recommended due the required accuracy of $\mathcal{L}_{i2}$. Even a small error when raised to the power of $N$ can spiral out of control.

\subsection{Log-space}

Following from the previous section, as our reweighting $\mathcal{L}_2$ is raised to the power of the number of our observations, they should definitely be computed in log-space, which turns the power into a linear factor. As most probabilistic work is already computed in log-space, this subsection barely needs to be stated. However, whilst working in log-space an efficient way of increasing the accuracy of the approximate analytic correction is to fit the correction such that the spread of the distribution $\log\left(\mathcal{L}_2\right)$ is minimised.

\section{Conclusion}
\label{sec:conclusion}

Sample selection is a pervasive issue in many scientific domains. For simple cases of sample selection which can be encapsulated with analytic functions, it is possible to analytically correct likelihood surfaces by introducing selection efficiency into the likelihood formulation. When analytic corrections fail to provide an adequate description of selection effects, Monte Carlo integration can be used on top of analytic approximations to further correct the likelihood surface, provided the experiment can be effectively simulated.

\section*{Acknowledgments}

We gratefully acknowledge the input of the researchers that were consulted during the creation of this paper, especially the Dark Energy Survey Supernova Working Group. Parts of this research were conducted by the Australian Research Council Centre of Excellence for All-sky Astrophysics (CAASTRO), through project number CE110001020.

\bibliography{bibliography}

\bsp	
\label{lastpage}
\end{document}